\documentclass[twocolumn,prl,nofootinbib,superscriptaddress]{revtex4}
\usepackage{graphicx}
\usepackage{amssymb}

\newcommand{\neb}{{\bar{\nu}_e}}

\newcommand{\nc}{\newcommand}
\nc{\beq}{\begin{equation}}   \nc{\eeq}{\end{equation}}
\nc{\bea}{\begin{eqnarray}}   \nc{\eea}{\end{eqnarray}}
\nc{\baa}{\begin{array}}      \nc{\eaa}{\end{array}}
\nc{\bit}{\begin{itemize}}    \nc{\eit}{\end{itemize}} 
\nc{\ben}{\begin{enumerate}}  \nc{\een}{\end{enumerate}}
\nc{\bce}{\begin{center}}     \nc{\ece}{\end{center}}
\def\beqa{\begin{eqnarray}}
\def\eeqa{\end{eqnarray}}
\def\be{\beq}
\def\ee{\eeq}
\def\to{\rightarrow}

\begin{document}

\title{Solar mass-varying neutrino oscillations}
\author{V. Barger} 
\author{Patrick Huber}
\affiliation{Department of Physics, University of Wisconsin, Madison, WI 53706}
\author{Danny Marfatia}
\affiliation{Department of Physics and Astronomy, University of Kansas, Lawrence, KS 66045}


\begin{abstract}
We propose
 that the solar neutrino deficit may be due to oscillations of
mass-varying neutrinos (MaVaNs). This scenario elucidates solar neutrino 
data beautifully while remaining comfortably compatible with 
atmospheric neutrino
and K2K data and with reactor antineutrino data at short and long baselines 
(from CHOOZ and KamLAND). We find that the survival 
probability of solar MaVaNs is independent of how
the suppression of neutrino mass caused by the acceleron-matter couplings
varies with density.
Measurements of 
MeV and 
lower energy solar neutrinos will provide a rigorous test of the idea.

\end{abstract}

\maketitle

\textbf{Introduction}.
Neutrino oscillation experiments have conclusively demonstrated that
neutrinos have mass. Also, evidence has mounted that the expansion of our
universe is in an accelerating phase caused by a negative pressure component
called dark energy. While these two seemingly disparate advances are
unequivocally among the most important of the last few years, our knowledge 
of both is woefully incomplete. 

Solar neutrino data provide the only evidence
of matter effects on neutrino oscillations. The Large Mixing Angle (LMA)
solution, while favored by reactor antineutrino data is somewhat discrepant
with solar data, thus calling into question how well neutrino-matter
interactions are understood. It is crucial that our understanding
of how neutrinos interact with matter be confirmed or modified.

Dark energy is troubling because the acceleration of the universe is a
very recent phenomenon in its expansion history. This 
``cosmic coincidence'' problem can be expressed as follows: 
Why are the dark matter and dark energy densities 
comparable today even though their ratio scales as $\sim 1/a^3$ (where
$a$ is the scale factor)?

The coincidence that the scale of dark energy 
$(2\times 10^{-3}$ eV)$^4$ is similar to the scale of neutrino mass-squared
differences (0.01 eV)$^2$ was exploited recently in 
Refs.~\cite{newmavans,mavans} to solve the coincidence problem. 
The authors of Ref.~\cite{mavans} considered the possibility of coupling
neutrinos to dark energy by supposing that the dark energy density is a
function of neutrino mass
 and imposing the condition that the total
energy density of neutrinos and dark energy remain stationary under
variations in neutrino mass. Then neutrino masses vary in such a way that
the neutrino energy density and the dark energy density are related over
a wide range of $a$.

A way to make the dark energy
density neutrino-mass-dependent is to introduce
a Yukawa coupling between a sterile neutrino and a
 light scalar field
(similar to quintessence)
called the acceleron. At energy scales
below the sterile neutrino mass, the effective potential of the acceleron
at late times receives a
contribution
equal to $m_\nu n_\nu$, where $m_\nu$ and $n_\nu$ are the active neutrino
mass and number density, respectively. Supersymmetric models
of neutrino dark energy have been constructed~\cite{super}.

Model-independent tests of neutrino dark energy are
cosmological~\cite{mavans,peccei}. 
A strict relationship between the dark energy
equation of state
and neutrino mass is predicted.
Further, since neutrino masses are predicted to scale with redshift
approximately as $(1+z)^{-3}$ in the nonrelativistic regime, 
cosmological and terrestrial probes of neutrino mass
could give conflicting results. 
If the acceleron
 couples both to neutrinos and matter, it may be
possible to investigate this scenario through neutrino 
oscillations~\cite{mavans2}. 
The coupling to matter is model-dependent.
The effective neutrino mass in matter is altered by the interactions
via the scalar which in turn modifies neutrino oscillations. 
 
For environments of 
approximately constant matter density, a satisfactory approach is
to parameterize the effects of the nonstandard interactions by effective 
masses and mixings in the medium~\cite{mavans2}. 
However, for solar neutrino oscillations it is not
possible to account for the exotic matter effects by a 
constant shift
in the oscillation parameters because the 
matter density in the sun varies by several orders of magnitude. 

In this letter we investigate solar MaVaN oscillations; these have not
been studied previously.
 We will show that 
since the neutrinos  propagate
adiabatically, the specific dependence of the evolving masses on the
acceleron potential is
irrelevant, so the predicted survival probability
depends only on the masses at their production sites. 
We then demonstrate how
MaVaNs improve the agreement with solar neutrino 
data while being 
perfectly consistent with KamLAND data~\cite{Barger:2003qi}. Finally, 
we illustrate via a calculation
of the survival probabilities of atmospheric muon neutrinos crossing the 
earth's core 
that the scheme is consistent with atmospheric neutrino data.
Since we focus on astrophysical and terrestrial neutrinos, 
the dependence of the neutrino mass on redshift 
is not pertinent to our considerations. 

\textbf{Effect of acceleron interactions on neutrino masses}.
At low redshifts, the contribution to the neutrino mass caused by the 
interactions of 
the acceleron with electrons and neutrinos can be written as
\begin{equation}
M_i=\lambda_{\nu_i}
(\lambda_e n_e+ \sum_i \lambda_{\nu_i} (n_{\nu_i}^{C\nu B} + 
{\frac{m_{\nu_i}}{E_{\nu_i}}} n_{\nu_i}^{rel}))/m^2_{\phi}\,,
\label{eq:m} 
\end{equation}
where $\lambda_{\nu_i}$ ($\lambda_e$) is the Yukawa coupling of the acceleron
to $\nu_i$ (the electron). Throughout,
when we quote values for $\lambda$, we mean $|\lambda|$.
In principle, the scalar $\phi$ has a mass, $m_{\phi}$, that 
depends on $n_e$ and the $n_{\nu_i}$. 
This dependence is weak since 
the underlying assumption in 
obtaining Eq.~(\ref{eq:m}) is that $\phi$ does not fluctuate signficantly
from its background value in the current epoch.
The number density of the cosmic neutrino
background in one generation of neutrinos and antineutrinos is 
 $n_{\nu_i}^{C\nu B} \sim 112$ cm$^{-3} \sim 10^{-12}$ eV$^3$, 
the number density of relativistic neutrinos in the background frame is
$n_{\nu_i}^{rel}$, and the electron number density is $n_e$. Here, 
$m_{\nu_i}$ are neutrino masses in a background dominated environment. 
We assume the heaviest ${\nu_i}$ to be ${\cal{O}}(0.05)$ eV in the present
epoch, and that as a result of their nonnegligible
velocities, the neutrino overdensity in the Milky Way
from gravitational clustering can be neglected~\cite{wong}.
 Then, $m_{\nu_i}$ essentially represent
 the masses of terrestrial neutrinos in laboratory experiments
like those measuring tritium beta decay. Since the neutrinos under
consideration are light, we do not expect the instabilities of
highly nonrelativistic neutrino dark energy~\cite{zald}. 

In principle, we should include a nucleon-acceleron Yukawa coupling. 
Since the electron-acceleron and nucleon-acceleron couplings are arbitrary
(within bounds from gravitational tests), we can parameterize
their combined effect on $M_i$ through $\lambda_e$, although
 this is not rigorously 
true for two reasons: (1) Conventional matter effects~\cite{msw} 
for active neutrino
oscillations do not depend on the nucleon number density $n_N$. (2) The $n_e$ 
and $n_N$ distributions in the sun do not have the same 
shape~\cite{bahcallrev}. Nonetheless,
this simplification suffices for our purposes.

Tests of the gravitational inverse square law require the coupling of
a scalar to the square of the gluon field strength to be smaller than
$0.01 m_N/M_{Pl} \sim 10^{-21}$~\cite{Adelberger:2003zx}, 
where $m_N$ is the nucleon mass. 
Since we have chosen to embody the effects of the couplings of the
acceleron to the nucleons and electrons in $\lambda_e$, the latter bound 
applies to $\lambda_e$.
In the region of the solar core where $pp$ neutrinos are produced, 
$n_e^0 \sim 60 N_A/$cm$^3 \sim 10^{11}$ eV$^3$~\cite{bahcallrev}. 
(Here and henceforth, we denote
the electron number density at the point of neutrino production by $n_e^0$). 
Thus, for $\lambda_e$ close to its upper bound, 
$\lambda_e n_e^0 \sim 10^{-10}$ eV$^3$. 


The cosmic neutrino background contributes negligibly to the mass shift
even for $\lambda_{\nu_i}$ of ${\cal{O}}(1)$. 
The $pp$ reaction creates neutrinos with the highest number density in the
production region ($\sim 7\times 10^{-8}$ eV$^3$) and 
lowest energies ($E_{\nu} \sim 0.3$ MeV) of all other processes in the $pp$
chain and $CNO$ cycle. Thus, $pp$ neutrinos have the highest
possible $m_\nu n_\nu^{rel}/E_\nu$, which for $m_\nu$ of ${\cal{O}}(1)$ 
eV is at most $n_{\nu_i}^{C\nu B}$.  
In sum, the dominant contribution to the
mass shift at the creation point arises from the $\lambda_e n_e$ term. 

We require that some $M_i$ be 
${\cal{O}}(10^{-3}-10^{-2})$ eV at neutrino production in the sun. Then,
for an assumed $m_\phi^2$ of ${\cal{O}}(10^{-11})$ eV$^2$, we need 
$\lambda_{\nu_i} \sim 10^{-4}-10^{-3}$.  
For this range of $\lambda_{\nu_i}$, 
the cosmic neutrino contribution in 
Eq.~(\ref{eq:m}) is five to six orders of magnitude 
smaller than the electron 
contribution, and the $pp$ neutrino contribution is eight to nine 
orders of magnitude smaller. The cosmic neutrino 
background density becomes dominant only after $n_e$ drops by about
six orders of magnitude. This does not happen until neutrinos reach 
the surface of the sun. As the neutrinos leave the sun,   
$m_\nu$ approaches its background value. The choice 
$\lambda_{\nu_i} \sim 10^{-3}$ serves more than one purpose. In addition
to fixing the maximum values of $M_i$, it ensures that $n_{\nu_i}^{C\nu B}$
 can be neglected for 
the entire path of the neutrinos through the sun. 

\textbf{Solar MaVaN oscillations}.
In the framework of the Standard Model (SM) with massive neutrinos and
 conventional
neutrino-matter interactions, solar (atmospheric) neutrinos oscillate with 
$\delta m^2_s \sim 8\times 10^{-5}$ eV$^2$ and 
$\theta_s \sim \pi/6$ ($|\delta m^2_a|\sim 0.002$ eV$^2$  
and $\theta_a \sim \pi/4$~\cite{Barger:2003qi}). In our
notation, $\delta m^2_s$ ($\delta m^2_a$) is the solar (atmospheric) 
mass-squared difference and $\theta_s$,
$\theta_a$ and $\theta_x$ are the mixing angles conventionally denoted 
by $\theta_{12}$, $\theta_{23}$ and $\theta_{13}$, 
respectively~\cite{Barger:2003qi}.
We also know that solar and atmospheric neutrino oscillations largely
occur independently of each other because $\theta_x$ must be small
from the nonobservance of $\neb \to \bar{\nu}_\mu$ oscillations at the 
atmospheric scale. 
In fact, data from the CHOOZ experiment demand 
$\sin^2 \theta_x \lesssim 0.05$ at the 
2$\sigma$ C.~L. in the conventional picture.

With the additional freedom that the $M_i$ provide, there is no reason to
believe that the three neutrino oscillation dynamics factorizes into the
dynamics of two two-neutrino subsystems. Nevertheless, since our purpose here
is to show that MaVaN oscillations are consistent 
with solar and atmospheric neutrino data while obeying the CHOOZ bound, 
we are entitled to accomplish
our goal via construction. A simplifying assumption is that the decoupling 
of solar and atmospheric neutrino oscillations continues to hold for MaVaNs. 
Then, the CHOOZ bound is automatically satisfied
and we need to demonstrate that the two neutrino framework 
is adequate for both neutrino anomalies. 

The evolution equations for solar MaVaN oscillations in the two-neutrino
framework are
\begin{eqnarray}
i{\frac{d}{dr}}  \left( \begin{array}{c} \nu_e \\ \nu_\mu
 \end{array} \right) 
&=&  \frac{1}{2E_\nu} \left[ U \left(\begin{array}{cc}(m_1-M_1(r))^2 & M_3(r)^2 \\
M_3(r)^2 & (m_2-M_2(r))^2
\end{array} \right) U^\dagger  \right. \nonumber\\
&& {} + \left.
\left( \begin{array}{cc} A(r) & 0 \\
0 & 0 
\end{array} \right) \right]
\left( \begin{array}{c} \nu_e \\ \nu_\mu \end{array} \right).
\label{evol} 
\end{eqnarray}
Here, $M_i$ is a linear combination of those in Eq.~(\ref{eq:m}), 
$U$ is the usual
$2\times2$ mixing matrix,
$E_\nu$ is the neutrino energy and
$
 A(r)=2 \sqrt{2} \,G_F n_e(r) E_\nu = 1.52 \times 10^{-7} {\rm{eV}}^2 \,
n_e(r) \, E_\nu \,(\rm{MeV})$
 is the 
amplitude for $\nu_e-e$ forward scattering in matter with $n_e$ in units
of $N_A/$cm$^3$. 
For typical $^8$B neutrinos ($E_\nu \sim 7$ MeV) $n_e^0 \simeq 100$, 
for $^7$Be neutrinos ($E_\nu \sim 0.9$ MeV) 
$n_e^0 \simeq 90$, and
for $pp$ neutrinos ($E_\nu \sim 0.3$ MeV) $n_e^0 \simeq 60$. The matter term
$A^0$ at the points of origin 
 is about $10^{-4}$ eV$^2$, 
$10^{-5}$ eV$^2$ and $10^{-6}$ eV$^2$
for $^8$B, $^7$Be and $pp$ neutrinos, respectively. With our choice of
 $|M_i^2|$ of ${\cal{O}}(10^{-5}-10^{-4})$ eV$^2$ at neutrino production,
we expect nonstandard matter effects to be of the same order 
as standard matter effects.

We adopt a matter dependence of the form,
\be
M_i(r)=\mu_i (n_e(r)/n_e^0)^k\,,
\label{k}
\ee
where
$k$ parameterizes the dependence of the neutrino mass on $n_e$, and
$\mu_i$ is the neutrino mass shift at the point of neutrino production.
As noted above we expect $k$ to be close
to unity, but we shall show that a wider range of $k$
is allowed by oscillation data.
We have implicitly
made the approximation that all neutrinos are created with the same 
values of $\mu_i$ irrespective of where in the sun they are produced.
Since almost all solar neutrinos are produced within $r<0.2 r_\odot$, for
which $n_e$ falls by about a factor of 3 from its value at the center of the 
sun, we consider the approximation to be reasonable.

We make the parameter choices 
$\mu_1=m_1=0$, $\mu_2=0.0077$ eV, $\mu_3=i 0.0022$ eV, $m_2=0.0089$ 
eV and $\theta=0.62$. The value of $\delta m^2$ in a background dominated
environment is $m_2^2=7.9\times 10^{-5}$ eV$^2$. We will sometimes refer
to $\mu_i$ as MaVaN parameters and $m_i$ as background parameters.
As we show, this set of parameters is consistent
with KamLAND data and improves the agreement with solar data. 

The evolution of the mass eigenstates as they travel
through the sun is governed by
\begin{eqnarray}  
4iE_\nu {\displaystyle\frac{d}{dr}} \left( \begin{array}{c} \nu_1 \\ \nu_2
 \end{array} \right) 
&=&  \left(\begin{array}{cc} -\Delta(r) & -4iE_\nu d\theta_m/dr \\
4iE_\nu d\theta_m/dr & \Delta(r)
\end{array} \right)
\left( \begin{array}{c} \nu_1 \\ \nu_2 \end{array} \right) \nonumber
\label{tevol} 
\end{eqnarray}
where $\Delta(r)$ is the magnitude of the mass-squared difference 
of the eigenvalues of the matrix in square brackets 
in Eq.~(\ref{evol}) and $\theta_m$ is 
the effective mixing angle in matter. The value of $\theta_m$ 
at the creation point of the neutrino is
\be
\cos 2\theta_m^0={\frac{-B-2\mu_3^2 \sin 2\theta}
{\sqrt{B^2+(m_2-\mu_2)^4 \sin^2 2\theta+
4 \mu_3^2 (A^0 \sin 2\theta +\mu_3^2)}}}\,,
\label{eq:thetam}
\ee
where $B=A^0-(m_2-\mu_2)^2 \cos 2\theta$,
which yields the standard result in the limit that 
$\mu_2, \mu_3 \to 0$. With
$
Q(r) = {\frac{\Delta(r)}{4E_\nu |d\theta_m/dr|}}\,,
$
the condition for adiabatic evolution~\cite{adiabatic} 
is $Q \gg 1$. 

\begin{figure}[t]
\centering\leavevmode
\mbox{\includegraphics[angle=270,width=8.7cm]{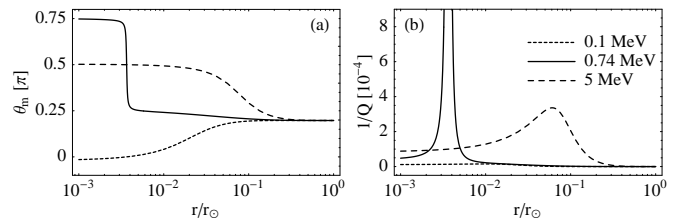}}
\caption[]{(a) $\theta_m$ and (b) $Q^{-1}$ as a function of
$r/r_\odot$ for three representative energies. 
The adiabatic condition $Q \gg 1$ is violated for 
$E_\nu=0.74$ MeV at $r/r_\odot \sim 0.0035$ because $d\theta_m/dr$ becomes 
very large. 
\label{fig:Q}}
\end{figure}

In Fig.~\ref{fig:Q}, we show how $\theta_m$ and $Q^{-1}$ 
depend on $r/r_{\odot}$
for $E_\nu=0.1, 0.74, 5$ MeV with $k=1$. 
We do not show the evolution of $\Delta$ since it is smooth throughout. 
Notice the step in $\theta_m$ at 
$r/r_\odot \sim 0.0035$ for $E_\nu=0.74$ MeV. (The energy at which this 
occurs depends on the background and MaVaN parameters chosen). 
The step manifests itself as a large spike in $Q^{-1}$; $Q < 10$  only  
in a 1 keV spread around 0.74 MeV. 
While adiabaticity is violently violated in this 
narrow range of energy, it is 
undetectable because \mbox{experimental resolutions are much larger than 
1 keV.}

For all practical purposes, the evolution is adiabatic and the
survival probability is given by the standard formula~\cite{parke},
$
P(\nu_e \to \nu_e) = (1+ \cos 2\theta_m^0 \cos 2\theta)/2\,,
$
with $\cos 2\theta_m^0$ from Eq.~(\ref{eq:thetam}).
Thus, we find that the survival probability of solar neutrinos is independent
of $k$ so long as the neutrinos propagate adiabatically. The dependence
on the acceleron-matter couplings enters only at the production point
of the neutrino via the $\mu_i$.

\begin{figure}[htb]
\centering\leavevmode
\mbox{\includegraphics[width=2.5in]{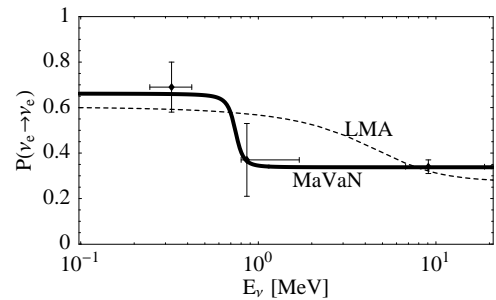}}
\caption[]{$P(\nu_e \to \nu_e)$ vs. $E_\nu$ for MaVaN
oscillations (solid curve). 
The dashed curve corresponds to conventional oscillations with the best-fit
 solution to KamLAND data. 
\label{fig:Pee}}
\end{figure}

\textbf{MaVaN oscillations vs.\ data}.
We now compare the predictions of this framework with solar data. To this end,
we use the recently extracted average survival probabilities of
the low energy ($pp$), intermediate energy ($^7$Be, $pep$, $^{15}$O, 
and $^{13}$N) and high energy ($^8$B and $hep$) neutrinos; for details
see Ref.~\cite{pee}.
From Fig.~\ref{fig:Pee}, 
we see that the MaVaN survival probability almost passes through the 
central values of the three data points. The agreement with intermediate 
energy data is 
remarkably improved compared to the LMA solution because
$P_{MaVaN}(\nu_e \to \nu_e)$ approaches $\sin^2\theta$ for lower 
$E_\nu$ than
for $P_{SM}(\nu_e \to \nu_e)$.
For the same solution, 
it is possible for $pp$ neutrinos to have a higher survival probability
than the vacuum value, $1-0.5 \sin^2 2\theta$.
This control over the width of the transition region and the larger
difference
between the survival probabilities of the $pp$ and $^8$B neutrinos 
(than $\cos^2 \theta \cos 2\theta$ for the LMA solution)
is a result of the freedom 
provided by the additional free parameter $\mu_3$.
Keeping in mind
that the survival probability of the neutrinos incident on earth is 
independent of $k$, if we set $\mu_3=k=0$,
we recover the standard MSW case with $m_2$ replaced by $m_2 - \mu_2$. The
$k$-dependence reappears for neutrinos passing through the earth.

\begin{figure}[t]
\centering\leavevmode
\mbox{\includegraphics[angle=270,width=3.2in]{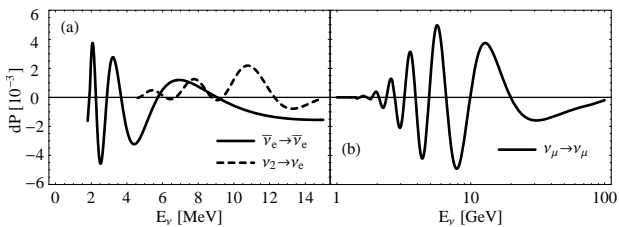}}
\caption[]{(a)  
$dP(\neb \to \neb)$
for reactor antineutrinos incident 
at KamLAND (solid), and
 $dP(\nu_2 \to \nu_e)$
for solar neutrinos passing through the center of the earth (dashed). 
(b) $dP(\nu_\mu \to \nu_\mu)$ 
for atmospheric neutrinos passing through
the earth's core. 
\label{fig:kamland}}
\end{figure}

An important question is whether MaVaN oscillations are consistent
with KamLAND data. The solid curve in Fig.~\ref{fig:kamland}a is
$dP(\neb \to \neb) \equiv P_{SM}(\neb \to \neb)-P_{MaVaN}(\neb \to \neb)$,
 for a mean KamLAND 
baseline of 180 km and energy resolution 7.3\%$/\sqrt{E(MeV)}$. Here, 
$P_{SM}$ 
is calculated for $k=1/2$ with $\delta m^2 = 8 \times 10^{-5}$ eV$^2$ 
and $\theta=0.55$; the latter
are the vacuum parameters favored by KamLAND data.
Since mass-varying effects in the
earth scale like the ratio of electron number density in the earth to that
in the sun, the effects are larger for {\it smaller} $k$.
Any $k > 1/2$ produces a $dP$ that is acceptable.

Another relevant
 question is if earth-matter effects are substantial 
for solar MaVaNs. Since $\nu_e$ with energy above a few MeV 
exit the sun and 
arrive at the earth in the intermediate neutrino mass eigenstate 
(denoted by
$\nu_2$), it is appropriate to study $\nu_2 \to \nu_e$ transitions to
assess the size of these effects.
The dashed curve in Fig.~\ref{fig:kamland}a is the energy-averaged
$dP(\nu_2 \to \nu_e)$
(assuming a 10\% energy resolution)
for neutrinos passing through the center of the earth in which case 
matter-effects are
expected to be enhanced.
$P_{MaVaN}(\nu_2 \to \nu_e)$ deviates only slightly from the usual 
matter oscillations
and is in accord with
a tiny
day-night effect as required by Super-Kamiokande and 
SNO data. 

We next show in Fig.~\ref{fig:kamland}b 
that atmospheric neutrino data are also consistent
with 2-neutrino $\nu_\mu \to \nu_\tau$ oscillations for MaVaN 
parameters of comparable size to those in the solar sector 
($k=1/2$, $\mu_2^a=0.01$ eV,
$\mu_3^a=0.003$ eV  and $m_3=0.047$ eV). 
Here, $P_{SM}$ is calculated for
$\delta m^2 = 0.0021$ eV$^2$ and $\theta=\pi/4$. 
$dP(\nu_\mu \to \nu_\mu)$ is averaged over the earth's core ($\cos \theta_Z =0.8-1$, where $\theta_Z$ is the nadir angle),
for a 10\% energy resolution. 
We also confirm that  $dP(\nu_\mu \to \nu_\mu)$ at the K2K
baseline is well below experimental sensitivity.

\textbf{Conclusions}.
We have shown that oscillations of variable mass neutrinos (that result
in exotic matter effects of the same size as standard matter effects) 
lead to an improved agreement (relative to conventional oscillations) 
with solar neutrino data while remaining compatible with KamLAND,
 CHOOZ, K2K and atmospheric data.

MaVaN oscillations are perfectly compatible with solar
data because the survival probability can change from a higher-than-vacuum 
value (at low energies)
to $\sin^2 \theta$ (at high energies) over a very narrow range of energies.
Since the neutrino propagation is highly adiabatic,
 the survival probability of solar neutrinos is 
independent of $k$ as defined in Eq.~(\ref{k}).

Whether or not an explanation of solar neutrino data requires MaVaN
oscillations will be answered by experiments that will measure 
 the survival probability of MeV and lower energy neutrinos.
As shown in Ref.~\cite{lsnd}, other tests in reactor and long-baseline 
experiments emerge when 
the scheme presented here is embedded
in a comprehensive 
model that can explain all extant neutrino oscillation data including
the LSND anomaly and a future
MiniBooNE result.

We have considered neutrinos with background mass of ${\cal{O}}(0.01)$ eV.
For such light neutrinos, only model-dependent (neutrino
oscillation) tests of the MaVaN scenario 
are viable because the model-independent (cosmological) tests become
inoperable.  There are two reasons for this: 
(1) The dark energy behaves almost exactly 
as a cosmological constant today. (2) If these light neutrinos do not 
cluster sufficiently, the local neutrino mass is the same as 
the background value, which is below the sensitivity
of tritium beta-decay
experiments. Then, high-redshift cosmological data (which should show
no evidence for neutrino mass) and data from 
tritium beta-decay experiments will be consistent.


\textbf{Acknowledgments}. We thank R.~Garisto, J.~Learned, 
S.~Pakvasa, A.~Smirnov, G.~Steigman and K.~Whisnant for helpful comments. 
We have made use of the
 GLoBES software~\cite{globes}.
This research was supported by the DOE under
Grant No. DE-FG02-95ER40896, by the NSF under Grant No. EPS-0236913,
by the State of Kansas \mbox{through KTEC and by the WARF.}




\end{document}